\documentclass[a4paper,12pt]{article}
\usepackage[margin=2cm]{geometry}
\usepackage{comment}
\usepackage{amsmath,amssymb,extarrows,graphicx,subfigure,setspace}
\usepackage{cite}
\usepackage{slashed}
\usepackage[toc,page]{appendix}
\usepackage{color}
\usepackage{hyperref}
\usepackage{dirtytalk}
\hypersetup{colorlinks=true, linkcolor=blue, citecolor=red, linktoc=page}
\makeatother

\usepackage{hyperref}

\newcommand{\be}{\begin{equation}}
\newcommand{\bea}{\begin{eqnarray}}
\newcommand{\eea}{\end{eqnarray}}
\newcommand{\ba}{\begin{array}}
\newcommand{\ea}{\end{array}}
\newcommand{\ee}{\end{equation}}
\newcommand{\bes}{\begin{equation*}}
\newcommand{\beas}{\begin{eqnarray*}}
\newcommand{\eeas}{\end{eqnarray*}}
\newcommand{\bas}{\begin{array*}}
\newcommand{\eas}{\end{array*}}
\newcommand{\ees}{\end{equation*}}

\numberwithin{equation}{section}
\begin{document}
	\onehalfspacing
	\noindent
	
	\begin{titlepage}
		\vspace{10mm} 
		
		
		\vspace*{20mm}
		\begin{center}
			
			{\Large {\bf Island in Warped AdS Black Holes}
			}
			
			\vspace*{15mm}
			\vspace*{1mm}
		{\bf \large Ankit Anand }
		\footnote{ E-mail : Anand@physics.iitm.ac.in}
		\vskip 0.5cm
		{\it
				Department of Physics, \\
				Indian Institute of
				Technology
				Madras, Chennai
				600 036, India}\\
			\vspace{0.2cm}

			\vspace*{1cm}
		\end{center}
		
\begin{abstract}

This paper investigates the Page curve in Warped Anti-de Sitter black holes using the \say{quantum extremal surface} prescription. The findings reveal that in the absence of an island, the entanglement entropy of Hawking radiation grows proportionally with time and becomes divergent at later times. However, when considering the island's emergence, which extends slightly beyond the event horizon, the growth of the entanglement entropy of Hawking radiation comes to a constant value. Eventually, the constant value is precisely twice the Bekenstein-Hawking entropy. We have also discussed the Page time as well as the Scrambling time.

\hspace{5 cm}\\
\hspace{5 cm}\\
\hspace{5 cm}\\
\hspace{5 cm}\\
\hspace{5 cm}\\
\hspace{5 cm}\\
\hspace{5 cm}\\
\hspace{5 cm}


\end{abstract}
\end{titlepage}

\newpage
\tableofcontents


\section{Introduction}
\label{Sec: Introduction}

\begin{figure}[ht]
	\begin{center}
		\includegraphics[scale=0.50]{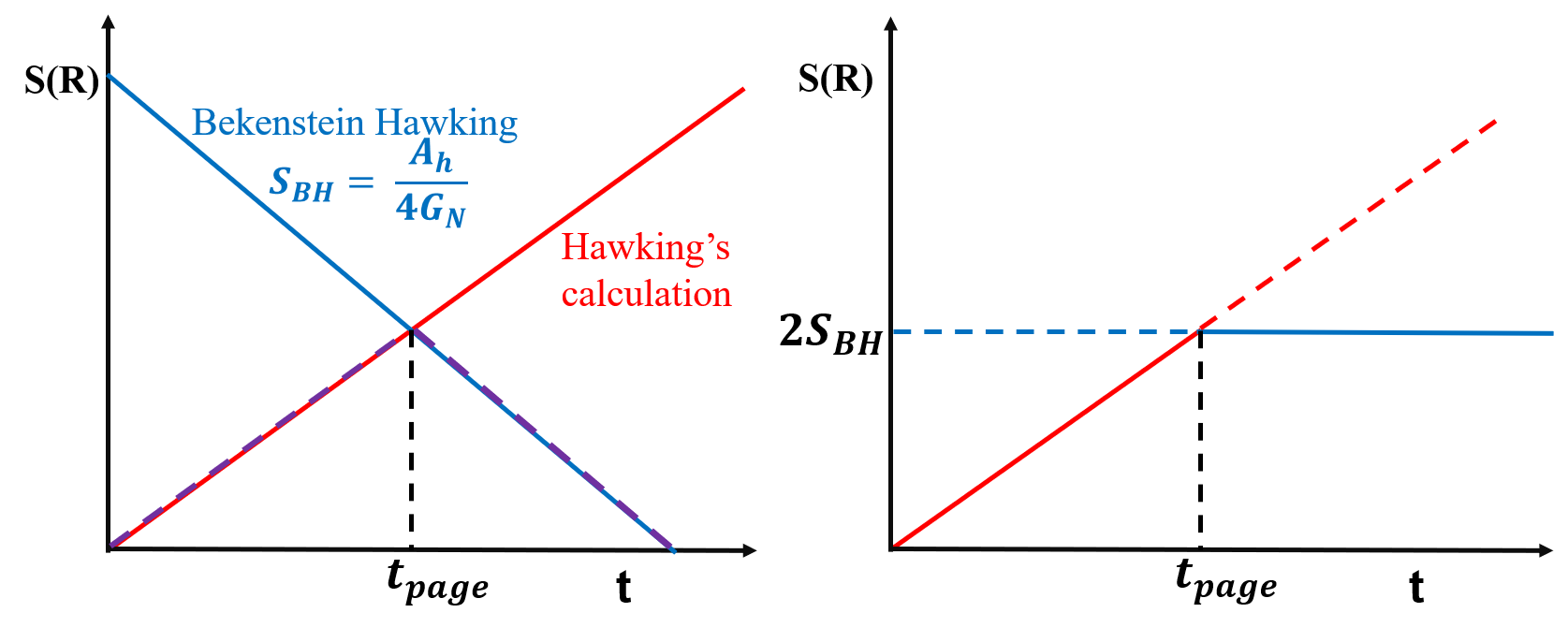}
	\end{center}
	\caption{\textbf{Left:}
	The Red Curve Shows the Hawking Radiation's Naïve Fine-Grained Entropy. It Grows Linearly in Time. The Blue Curve Shows the Thermodynamic Calculation of Bekenstein-Hawking Entropy. The Purple Dotted Line Shows the Actual Page Curve for an Evaporating Black Hole. Here $A_h$ Denotes the Horizon Area, and $G_N$ is Newton's Constant and \textbf{Right:} 
	The red Curve Shows Hawking Radiation's Naïve Fine-Grained Entropy. It Grows Linearly in Time. The Blue Curve Shows the Inclusion of Island Makes a Constant Value of $S(R)$ as $2S_{BH}$.}
	\label{fig: Page curve}
\end{figure}

The black hole information paradox has been disputed for more than 40 years since Stephen Hawking observed that information may be lost during black hole evaporation \cite{Hawking:1976ra}.
It is believed that the resolution of the Information paradox will help in formulating the theory of Quantum Gravity. The black hole information problem was first brought to light by Hawking in 1975 when he proposed that Hawking radiation behaves like typical thermal radiations \cite{Hawking:1975vcx}. According to his findings, any information that falls into a black hole would seemingly vanish forever as the black hole evaporates. However, this conclusion contradicts a fundamental tenet of quantum mechanics – the unitarity principle. Suppose we assume that a black hole undergoes evaporation from a pure quantum state based on the principle of unitarity. In that case, it must ultimately end up in another pure quantum state instead of a mixed state. The Page curve depicts the behavior of the entropy of Hawking radiation during this process \cite{Page:1993wv, Page:2013dx}. Hence, whether the black hole information problem can be resolved hinges on whether the Page curve of an evaporating black hole (as seen from Figure Left \ref{fig: Page curve}) can be accurately reproduced.

By using the idea of the AdS/CFT correspondence and having a theoretically controlled theory of quantum gravity, one should explore black holes in (asymptotically) anti-de Sitter (AdS) space, such that one has a non-perturbative formulation of quantum gravity. Black holes in Anti-de Sitter space don't evaporate the page curve that these types of BH should follow, as seen from Figure \ref{fig: Page curve}. For a manageable model of an evaporating BH in Ads, authors of \cite{Penington:2019npb, Almheiri:2019psf} use the idea that AdS is coupled with an external bath such that it helps radiation to reach the boundary. Black holes can entirely evaporate and vanish in this manner. This results in Hawking radiation as a complete thermal spectrum. Consider a scenario in which a pure state collapses to create a black hole that produces thermal Hawking radiation. Then, because the unitarity principle states that the end state should be pure rather than mixed (thermal), it results in the famous black hole information paradox. 

\par 
 From the viewpoint of entropy, this contradiction may also be understood. It was widely assumed in the context of black hole physics that to explain the Page curve, and one may require a more profound knowledge of the microscopic description of black hole degrees of freedom. As we know, the entropy of a black hole generated from a pure state is zero. Due to Hawking radiation, the black hole would eventually disappear. Hawking radiation's entanglement entropy should rise with time as more Hawking quanta are produced and entangled with the black hole during the early stages of black hole evaporation. On the other hand, the black hole's thermodynamic or Bekenstein-Hawking entropy would drop as its horizon area shrank. When the Bekenstein constraint is violated, Hawking radiation's entanglement entropy behavior changes. The Bekenstein bound indicates that the black hole's fine-grained entropy should not exceed its Bekenstein-Hawking entropy. Because the residual black hole's degrees of freedom and the outgoing radiations are pure, the Hawking radiation's entanglement entropy is the same as the fine-grained entropy of the residual black hole. As a result, at the Page time, when the entanglement entropy of Hawking radiation equals the Bekenstein-Hawking entropy of the black hole, the Hawking radiation's entanglement entropy becomes constant by the inclusion of island as shown in Figure Right \ref{fig: Page curve}. This suggests that the original black hole's quantum information is encoded in Hawking radiation, and black hole evaporation is consistent with unitarity evolution. Thus, the unitarity evolution of black hole evaporation corresponds to the Page curve: the Hawking radiation's entanglement entropy grows monotonically in the early period of black hole evaporation, then achieves a maximum value at the Page time and eventually needs to fall to zero after the evaporation process. The Page curve is believed to be produced in a quantum gravity theory. So, not only for the resolution of the information loss paradox, but the theory of quantum gravity needs the reproduction of the Page curve. Nonetheless, it was recently demonstrated that at least in two dimensions, the Page curve might be explained within the semiclassical description of gravity \cite{Penington:2019npb, Almheiri:2019psf, Almheiri:2019hni, Almheiri:2019yqk}.

\par

Indeed, driven by holographic entanglement entropy \cite{Ryu:2006bv, Hubeny:2007xt} and the introduction of the quantum extremal surface \cite{Engelhardt:2014gca}, \cite{Almheiri:2019hni} proposes a new method for calculating fine-grained black hole entropy, based on which to access the entropy of the radiation, one should additionally include a hypothetical contribution of an island comprising a portion of the black hole interior. The von Newman entropy of Hawking radiation will rise monotonically, according to Hawking's calculations. Quantum theory, on the other hand, necessitates unitary black hole evaporation; therefore, the von Newman entropy of Hawking radiation should obey the Page curve provided information is retained during evaporation \cite{Page:1993wv}. The island proposal \cite{Almheiri:2019hni} provides a formula for the fine-grained entropy of Hawking radiation and is written as  
\begin{equation} \label{S(R)}
    S(R) = \text{min} \left\{\text{ext}\left[\frac{\text{Area}(\partial I)}{4G_N}+S_{\text{matter}}(I \cup R)\right]\right\} \ ,
\end{equation}
where $R$ is a region called the Radiation region, and it is far from the black hole where distant observers collect the radiation, and $I$ is a spatially disconnected island surrounding the horizon that is entangled with $R$. The idea behind this statement is that when about half of the black hole has disappeared, the Hawking radiation (about $I$) starts purifying the radiation released earlier (roughly $R$). This purification of early radiation by late Hawking radiation reflects the entanglement between the two portions and arises from the view of Hawking radiation as the result of the generation of entangled particle pairs near the horizon(This is interpreted as a vacuum). The slowly evaporating black hole has a decreasing area. Hence, S(R) reduces in time. As a result, $R \cup I$ purifies over time, and its entanglement does not change.

\par
The Island formula for JT black holes can be obtained from the gravitational path integral using the replica trick, as demonstrated in \cite{Penington:2019kki, Almheiri:2019qdq}. The authors explained that the Page curve of eternal black holes can be derived as follows: there are two saddles - the disconnected and connected saddles. Below the Page time, the disconnected saddle dominates, while after the Page time, the connected saddle (replica wormhole) becomes dominant. The disconnected saddle is responsible for the linear time growth of entanglement entropy in Hawking radiation, while the connected saddle contributes a finite amount. By combining the contributions from the disconnected and connected saddles to Hawking radiation's entanglement entropy, one can accurately reproduce the Page curve. It's worth noting that the argument presented in reference \cite{Penington:2019kki} also applies to \say{$n$} boundary wormholes, known as replica wormholes.

\par
Although the quantum extremal surface prescription was first given in the context of holography \cite{Ryu:2006bv, Hubeny:2007xt}, the island rule may be applied to black holes in more general theories. For various two-dimensional models, the formula was validated \cite{Almheiri:2019hni, Almheiri:2019yqk}. The island rule for two-dimensional gravity has been developed using the replica trick \cite{Calabrese:2009qy, Holzhey:1994we, Callan:1994py}, and the island contribution has been linked to replica wormholes \cite{Penington:2019kki, Almheiri:2019qdq}. In JT gravity, the Page curve for evaporating black holes has also been explored \cite{Hollowood:2020cou}. In \cite{Ageev:2019xii}, the wormhole configurations in JT gravity are studied. See \cite{Chen:2020jvn, Geng:2020fxl, Chen:2020uac, Chen:2020hmv, Hernandez:2020nem, Colin-Ellerin:2021jev, Colin-Ellerin:2020mva} for further development in the low-dimensional situation, as well as references therein, especially \cite{Gautason:2020tmk, Anegawa:2020ezn, Hartman:2020swn} for asymptotically flat two-dimensional spacetime. \cite{Chen:2019uhq, Hashimoto:2020cas, Wang:2021woy, Matsuo:2020ypv, Akal:2020twv, Almheiri:2020cfm, Dong:2020uxp, Balasubramanian:2020xqf, Raju:2020smc, Alishahiha:2020qza, Saha:2021ohr, Yu:2021rfg, Azarnia:2022kmp, Yadav:2022fmo, Ahn:2021chg, Basu:2022reu, Afrasiar:2022ebi, Omidi:2021opl, Anand:2022mla} looked at black holes in asymptotically flat spacetime in four and higher dimensions.

\par
For asymptotically AdS spacetimes, quantum information has been researched rather thoroughly; however, our understanding of other spacetimes, including asymptotically flat or de Sitter spacetimes, is far more restricted since we currently know very little about the dual field theory, assuming it even exists. The Warped $AdS_3/CFT_2$ correspondence \cite{Anninos:2008fx, Detournay:2012pc, Hofman:2014loa, Jensen:2017tnb} is an intriguing ultraviolet deformation of AdS/CFT where we have a lot of knowledge about the structure of the field theory dual.  It is a duality between gravitational theories in $2 + 1$ dimensions in a space with Warped AdS asymptotic and a hypothesized class of non-relativistic theories in $1 + 1$ dimensions known as warped conformal field theories (WCFTs), whose symmetry content includes copies of the Virasoro and the $U(1)$ Kac-Moody current algebras. Recent developments have helped to strengthen this duality; for instance, an analogue of the Cardy formula was developed in \cite{Detournay:2012pc}. Several authors, including \cite{Anninos:2013nja,Castro:2015csg,Azeyanagi:2018har,Song:2016pwx,Song:2016gtd}, explored the topic of entanglement entropy.

\par

Although the island structure and the Page curve are studied with other black hole geometries in higher dimensions, most of them, to the best of our knowledge., are considered black holes in asymptotically flat/AdS/dS. As a result, it is crucial to investigate if the island formula may be applied to other cases(non-asymptotically flat/AdS/dS), also known as nonstandard black hole geometries. Checking the Page curve for nonstandard black hole geometries using the island formula is critical for the island method's breadth of application and quantum gravity. We take a step forward in this approach in this work. We have studied the page curve in the Warped AdS black hole case.

\par 

The paper is organized as follows. In Section \ref{Sec: Review of the Warped AdS black holes}, we briefly introduce the Warped AdS black holes. Moving to Section \ref{Sec: Page Curve For Non-Rotating WAdS Black Hole}, we calculate the entanglement entropy without an island and demonstrate the information paradox for Warped AdS black holes. In Section \ref{Sec: With Islands}, we examine the behavior of generalized entropy during both early and late times with Island. Upon considering the construction of a single island, we observe the reproduction of the unitary Page curve. Section \ref{Sec: Page time and scrambling time} discusses the Page time and scrambling time. The last section presents our conclusion and discussion. Additionally, we present the isometries in Appendix \ref{Isometries of WAdS3}.

\section{Review of the Warped AdS black holes}
\label{Sec: Review of the Warped AdS black holes}

This section briefly reviews Warped AdS black holes based on \cite{Anninos:2008fx}. The action for three-dimensional topologically massive gravity (TMG) with a negative cosmological is as follows
\begin{eqnarray}
S_{\rm TMG} \!\!\!\! &=& \!\!\!\! \frac{1}{2\mathbb{K}}\int _\mathcal{M} d^3x \sqrt{-g} \left(R+\frac{2}{l^2}\right) \nonumber \\
&&+ \frac{l}{12 \nu \mathbb{K} } \int _\mathcal{M} d^3x \sqrt{-g} \left(\epsilon^{\alpha \beta \gamma} \Gamma^{\tau}_{\alpha \sigma}\left \{\partial_\beta \Gamma^{\sigma}_{\tau \gamma} +\frac{2}{3} \;\;\Gamma^{\sigma}_{\beta \rho} \Gamma^{\rho}_{ \tau \delta}\right\}\right) \  ,
\label{Action}
\end{eqnarray}
where the $\epsilon^{\alpha \beta \gamma}$ is the Levi-Civita tensor, $\nu$ is a dimensionless coupling constant called warp factor, $\mathbb{K}=8 \pi G_N$ and $G_N$ is Newton's constant. The wrap factor $\nu$ is related to the Graviton mass $\mu$ with the inclusion in the Chern-Simons action coefficient by $\nu=\mu l/3.$ At $\mu l = 1$ or $\nu = 1/3$, this is a critical chiral gravity theory for asymptotically AdS spacetimes. The equation of motion is obtained as follows
\begin{equation}\label{eom}
G_{\alpha \beta} = \frac{1}{l^2} g_{\alpha \beta} - \frac{l}{3 \nu} \left[ \epsilon_\alpha ^{\delta \tau} \nabla_\delta \left(R_{\tau \beta}-\frac{1}{4}g_{\tau \beta}R \right) \right] \ ,
\end{equation}
where $G_{\alpha \beta}$ is the Einstein tensor. We can show from \eqref{eom} that every Einstein vacuum solution is also a TMG solution because $G_{\alpha \beta}=g_{\alpha \beta}/l^2$. There exist non-Einstein solutions to these equations of motion, which will be called Warped AdS because a warped fibration is involved. To get the general solution, remember that AdS may be expressed as a Hopf fibration over Lorentzian (or Euclidean) $AdS_2$ with the real line as the fiber \cite{Duff:1998cr, Bengtsson:2005zj}. The metric may be expressed as a spacelike fibration with $u$ as the fibre coordinate
\begin{eqnarray} \label{Metric}
ds^2 = \frac{l^2}{4} \left[-\cosh^2\;\sigma d\tau^2\;+\;d \sigma^2\;+\;\left(du+\sinh\sigma d\tau \right)^2\right]
\end{eqnarray}
with $ -\infty < (u, \tau, \sigma) < \infty$. They have the isometries $SL(2, \mathbb{R})_L \times SL(2, \mathbb{R})_R$. We must multiply the fiber by a warp factor to generate warped AdS, which breaks the isometry group to $SL(2, \mathbb{R}) \times U(1)$, where $U(1)$ is non-compact. For a Spacelike warped AdS solution, we have to warp the metric \eqref{Metric} in the form
\begin{equation}\label{WADS-spacelike}
ds^2 = \frac{l^2}{\nu^2+3} \left[-\cosh^2(\sigma) d\tau^2+d \sigma^2+ \frac{4\nu^2}{\nu^2+3}(du+\sinh(\sigma) d\tau)^2\right] \ .
\end{equation}
We get the spacelike stretched AdS when $\nu^2 > 1$ and the spacelike squished AdS if $\nu^2<1$. $U(1)_L \times SL(2, \mathbb{R})_R$ gives the isometries\footnote{More detailed discussion can be found in appendix \ref{Isometries of WAdS3}} maintained by \eqref{WADS-spacelike}. At a value of $\nu^2=1$, the solution is not warped. Now we'll focus on asymptotic black hole solutions to warped AdS. Until now, only the asymptotically spacelike extended (i.e., $\nu^2>1$ ) situation \cite{Bouchareb:2007yx} has been known to have solutions devoid of naked CTCs or other pathologies. In Schwarzschild coordinates, the metric that describes spacelike stretched black holes is \cite{Moussa:2003fc, Anninos:2008fx}
\begin{equation}\label{WADS-BH-rotating}
\frac{ds^2}{l^2} = dt^2+\frac{dr^2}{(\nu^2+3)(r-r_+)(r-r_-)} + \left(2 \nu r -\sqrt{r_+r_-(\nu^2+3)}\right)dt d \phi + \frac{r}{4}\Xi(r) d \phi^2 \ ,
\end{equation}
where
\begin{equation}
\Xi(r) = 3(\nu^2-1)r+(\nu^2+3)(r_++r_-)-4 \nu \sqrt{r_+r_-(\nu^2+3)} \ ,
\label{Xi}
\end{equation}
and the coordinates $ 0<r<\infty \;,\; -\infty < t < \infty\;,\; \phi \sim \phi+2 \pi$. There are two horizons at $r_\pm$. For $\nu <1$, the solution has closed timelike curves. However, the solution is well-behaved for $\nu^2 \geq 1$. Therefore, we restrict ourselves to the case $\nu \geq 1$. This solution is a discrete quotient of the Warped AdS,
\footnote{There is a coordinate transformation that maps \eqref{WADS-BH-rotating} to \eqref{WADS-spacelike} \cite{Anninos:2008fx}.}
Similarly, a BTZ black hole is a discrete quotient of $AdS_3$ \cite{Anninos:2008fx}. Moreover, for $\nu = 1$, the metric \eqref{WADS-BH-rotating} reduces to a rotating Banados-Teitelboim-Zanelli  (BTZ) black hole \cite{Banados:1992wn,Banados:1992gq}. In this case, by applying the following coordinate transformations \cite{Auzzi:2018zdu} 
\begin{eqnarray} 
r &=& \tilde{r}^2 \;\;\;\;\;;\;\;\;\;\; t = \frac{\sqrt{r_+}-\sqrt{r_-} }{l^2}\;\;\;\;\;;\;\;\;\;\;\phi = \frac{l \tilde{\phi}-\tilde{t}}{l^2 (\sqrt{r_+}-\sqrt{r_-}) } \tilde{t} \;\;\;\;\;;\;\;\;\;\;r_\pm = \tilde{r}_\pm^2 \ . \nonumber 
\end{eqnarray}
one can rewrite the metric \eqref{WADS-BH-rotating} as follows
\begin{equation}
ds^2 = -\frac{(\tilde{r}^2-\tilde{r}_+^2-\tilde{r}_-^2)}{l^2} d\tilde{t}^2+\frac{l^2 \tilde{r}^2}{(\tilde{r}^2-\tilde{r}_+^2)(\tilde{r}^2-\tilde{r}_-^2)} d\tilde{r}^2 - 2 \frac{\tilde{r}_+ \tilde{r}_-}{l} d\tilde{\phi}d\tilde{t}+\tilde{r}^2 d \tilde{\phi}^2,
\end{equation} 
which is the metric of a rotating BTZ black hole.

\par

It is believed that TMG is dual to WCFT. In the holographic dictionary, the warping parameter $\nu$ is linked to the left and right central charges of the boundary WCFT, which are given by \cite{Anninos:2008qb}
\begin{equation}
C_L = C_R = \frac{12 l \nu^2}{G(\nu^2+3)^{3/2}} \ .
\end{equation}
The mass, thermal entropy, angular momentum, the temperature and the angular velocity of horizon are given by \cite{Anninos:2008fx}

\begin{eqnarray} \label{M-S-J}
    M   \!\!\!\! &=& \!\!\!\! \frac{(\nu^2 +3)}{16 G_N}  \left( (r_+ + r_-)  - \frac{\sqrt{r_+ r_- ( \nu^2 +3)}}{\nu} \right) \;\;\;;\;\;\; S  = \frac{ \pi l}{4 G_N} \left( 2 \nu r_+ - \sqrt{r_+ r_- (\nu^2 +3) }\right) \ ,
\cr && \cr
J  \!\!\!\! &=& \!\!\!\! \frac{l (\nu^2 +3) }{32 G_N} \left( \frac{r_+ r_- (5 \nu^2 +3)}{2 \nu } - (r_+ + r_-) \sqrt{r_+ r_- ( \nu^2 +3 )}\right) \ ,
\cr && \cr
T  \!\!\!\! &=& \!\!\!\! \frac{\nu^2+3}{4 \pi l} \frac{r_+-r_-}{2 \nu r_+-\sqrt{(\nu^2+3)r_+r_-}}\;\;\;\;\;;\;\;\;\;\;
\Omega= \frac{2}{(2 \nu r_+-\sqrt{(\nu^2+3)r_+r_-})l} \ .
\label{temperature}
\end{eqnarray}
\\The metric in \eqref{WADS-BH-rotating} can be written in Arnowitt-Desser-Misner (ADM) form as
\begin{equation}\label{wrapmericADm}
ds^2= - N(r)^2 dt^2 +\frac{l^4 dr^2}{4 R(r)^2 N(r)^2} + l^2 R(r)^2 \left(d \phi + N^\phi(r) dt\right)^2 \ ,
\end{equation}
where
\begin{equation*}
R(r)^2 =  \frac{r}{4} \Xi(r)\;\;\;;\;\;\;N(r)^2 = \frac{l^2(\nu^2+3)(r-r_+)(r-r_-)}{4R(r)^2}\;\;\;;\;\;\;
N^\phi(r) = \frac{2 \nu r- \sqrt{r_+r_-(\nu^2+3)}}{2R(r)^2} \ .
\end{equation*}
\subsection*{The Non-Rotating Black Hole}
The metric of the non-rotating black hole can be obtained by setting $r_- =0$ and $r_+ =r_h$ in \eqref{wrapmericADm}. The metric can be written as
\begin{equation}
ds^2= - N(r)^2 dt^2 +\frac{l^4 dr^2}{4 R(r)^2 N(r)^2} + l^2 R(r)^2 \left(d \phi + N^\phi(r) dt\right)^2 \ ,
\label{metric-non-rotating-ADM}
\end{equation}
where
\begin{equation}
R(r)^2=\frac{r}{4} \Xi(r) \ , \hspace{0.5cm} N(r)^2 = \frac{l^2(\nu^2+3)(r-r_h) r}{4R(r)^2}\ , \hspace{0.5cm} N^\phi(r)= \frac{ \nu r}{R(r)^2} \ ,
\end{equation}
\begin{equation}
\Xi(r) = 3(\nu^2-1)r+ r_h (\nu^2+3) \ .
\end{equation}

The area of the Event horizon can be computed with the relation
\begin{equation}
    A= 2 \pi R(r_h)= 2 \pi \nu r_h \ ,
\end{equation}
and in terms of area term, the black hole entropy can be written as 
\begin{equation}
    S_{\rm BH} = \frac{A}{4G_N} = \frac{\pi \nu r_h}{2 G_N}
\end{equation}
 We define a new angular coordinate $\psi$ as follows which is spacelike everywhere (See also \cite{Azarnia:2021uch,Azarnia:2022kmp,Yu:2021rfg})
\bea
\psi = \phi + N^{\phi}(r_h) t = \phi + \frac{\nu r_h}{R(r_h)^2} t,
\label{psi}
\eea 
Then the metric in \eqref{wrapmericADm} can be written as follows
\bea
ds^2= - N(r)^2 dt^2 +\frac{l^4 dr^2}{4 R(r)^2 N(r)^2} + l^2 R(r)^2 \left(d \psi + ( N^\phi(r) - N^\phi(r_h) )dt \right)^2 \ .
\label{metric-non-rotating-ADM-psi}
\eea

  At $\psi = \rm const$, the metric \eqref{metric-non-rotating-ADM-psi} becomes

\begin{eqnarray}
    ds^2 \!\!\!\! &=& \!\!\!\! - N(r)^2 dt^2 +\frac{l^4 dr^2}{4 R(r)^2 N(r)^2} + l^2 R(r)^2 \left( N^\phi(r) - N^\phi(r_h) \right)^2 dt^2 \ .
\end{eqnarray}
  By defining $\left( N(r)^2 - l^2 R(r)^2 \left( N^\phi(r) - N^\phi(r_h) \right)^2 \right) = \mathcal{N_R}(r)$ the above expression reduces to 
  \begin{eqnarray}\label{metric-non-rotating-ADM-psi-1}
\!\!\!\! &=& \!\!\!\! \left(\mathcal{N_R}(r)^2\right) \left( - dt^2 + \frac{l^4 dr^2}{4 R(r)^2 N(r)^2  \mathcal{N_R}(r)^2 }\right)
\cr && \cr
\!\!\!\! &=& \!\!\!\! \left(\mathcal{N_R}(r)^2 \right) \left( - dt^2 + dr^*(r)^2 \right),
  \end{eqnarray}

\begin{figure}
	\begin{center}
		\includegraphics[scale=0.65]{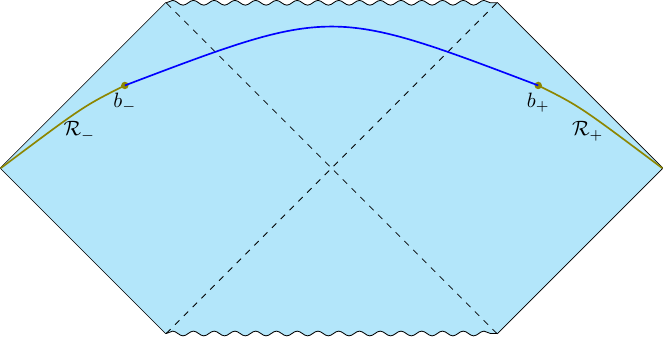}
		\hspace{1cm}
		\includegraphics[scale=0.65]{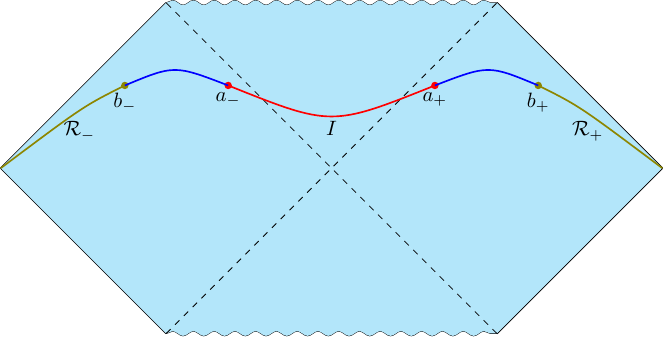}
	\end{center}
	\caption{
		The Penrose diagram of a Non-rotating WAdS black hole \cite{Jugeau:2010nq}. {\it Left}) When there are no Islands. {\it Right}) When an Island $\mathcal{I}$ is indicated in Red with endpoints $a_-$ and $a_+$. The radiation region is $\mathcal{R} = \mathcal{R}_+ \cup \mathcal{R}_-$ and indicated in Olive. The endpoints of $\mathcal{R}$ are denoted by $b_{\pm}$ and are located in the Baths. We assume that the state on the Full Cauchy slice is pure. Therefore, one can calculate the Entanglement Entropy of matter fields $S_{\rm matter} (\mathcal{R} \cup \mathcal{I})$ on the complement intervals $[b_-,b_+]$ in the left panel and $[b_-, a_-] \cup [a_+,b_+]$ in the right panel which is shown in Blue.}
	\label{fig: Penrose}
\end{figure}

From which one can find the tortoise coordinate as follows
\bea
r^*(r) \!\!\!\! &=& \!\!\!\! \int \frac{l^2 dr}{2 R(r) N(r) \mathcal{N_R}(r) }
\cr && \cr
\!\!\!\! & = & \!\!\!\! - \frac{4 \nu}{\nu^2 +3} \tanh^{-1} \sqrt{\frac{r (\nu^2 +3) }{ 3 r (1 -\nu^2) + 4 r_h \nu^2 }}
\cr && \cr
\!\!\!\! & = & \!\!\!\! \frac{2 \nu}{\nu^2 +3} \log \left( \frac{ \sqrt{ 3 r (1 -\nu^2) + 4 r_h \nu^2 } - \sqrt{r (\nu^2 +3)}}{ \sqrt{ 3 r (1 -\nu^2) + 4 r_h \nu^2 } + \sqrt{r (\nu^2 +3)}}\right)
\label{tortoise-non-rotating-new}
\eea 
In this case, the Penrose diagram is the same as that of a Schwarzschild black hole \cite{Jugeau:2010nq} (See Figure \ref{fig: Penrose}).
In the Kruskal coordinates
\begin{equation}\label{Kruskal-BB-new}
    U = - e^{- \kappa \left( t -r^*(r)\right)} \;\;\;\;\;\;\;\;\;\;;\;\;\;\;\;\;\;\;\; V=  + e^{ \kappa \left( t + r^*(r)\right)},
\end{equation}

By applying \eqref{tortoise-non-rotating-new} and \eqref{Kruskal-BB-new}, the metric at $\psi= \text{const}$ is written as follows

\begin{equation}\label{metric-black brane-Kruskal-1-new}
    ds^2 = - \frac{dU dV}{\mathcal{W}(r)^2} \ ,
\end{equation}
where 
\begin{eqnarray}\label{metric-black brane-Kruskal-1-new W(r)}
      \mathcal{W}(r) = \frac{2 \kappa r_h  \nu \left[\sqrt{4 r_h \nu^2-3(\nu^2-1)r}-\sqrt{(3+\nu^2)r}\right]}{\sqrt{(r-r_h)(4 r_h \nu^2-3(\nu^2-1)r)}\left[\sqrt{4 r_h \nu^2-3(\nu^2-1)r}+\sqrt{(3+\nu^2)r}\right]}
\end{eqnarray}

\section{Page Curve For Non-Rotating Warped AdS Black Hole}
\label{Sec: Page Curve For Non-Rotating WAdS Black Hole}


In this section, we evaluate the entanglement entropy of Hawking radiation emitted by a non-rotating Warped AdS black hole. The entanglement entropy is associated with the matter sector coupled with gravity. We will analyze the contributions to the entanglement entropy from configurations with and without islands, which are regions in spacetime crucial for the behavior of the entropy. We aim to understand the Page curve, which describes the evolution of entanglement entropy during the black hole evaporation process, and how it aligns with the unitarity principle of quantum mechanics.

\par

To simplify the calculation and without sacrificing generality, we will specifically examine the case of having one island present. We will demonstrate that considering this single-island configuration is sufficient to achieve finite entanglement entropy later, leading to a Page curve that aligns with the unitarity principle. We need to make certain assumptions for the subsequent calculation:

\begin{itemize}
 \item To disregard the influence of matter on spacetime, we assume that the radiation region is sufficiently distant from the region occupied by the black hole.

 \item To apply the s-wave approximation, we also assume that the separation between the points in the radiation region, denoted as $b_+$ and $b_-$, is significantly large.

 \item Initially, the system exists in a pure state, resulting in an equivalence between the entanglement entropy values in the radiation region $\mathcal{R}_+ \cup \mathcal{R}_-$ and the complementary region $[b_-, b_+]$.

 \item Let's consider two points, denoted as $d_+$ and $d_-$, with their coordinates in the $(t, r)$ space represented as $(t_d, d)$ and $(-t_d + \frac{i \beta}{2}, d)$ respectively. Here, $\beta$ represents the inverse of the Hawking temperature of the black hole.

\end{itemize}

Now, we calculate the Page curve for the metric \eqref{metric-non-rotating-ADM}. To absorb the UV divergent term in the Entanglement Entropy of the matter CFT (Conformal Field Theory), we renormalize Newton's constant as follows \cite{Susskind:1994sm, Almheiri:2020cfm, Hashimoto:2020cas, Alishahiha:2020qza, Azarnia:2021uch}
\begin{equation}\label{G_N-r-CFT}
    \frac{1}{4 G_{ N,r}} = \frac{1}{4 G_N} - \frac{1}{\epsilon} \ ,
\end{equation}
 
The endpoints $b_{\pm}$ of the left and right radiation regions and the Island $a_{\pm}$ are given as follows
\begin{eqnarray}\label{a,b}
    && a_+ :( t_a, a) ,\;\;\;\;\;\;\;\;\;\;\;\;  a_-: ( - t_a + i \beta /2, a) \ , 
    \cr && \cr 
&& b_+ :(t_b, b), \;\;\;\;\;\;\;\;\;\;\;\;  b_-: (- t_b + i \beta /2, b) \ .
\end{eqnarray}

We can write the distance $d(m_1,m_2)$ between two arbitrary points $m_1$ and $m_2$ in the bulk spacetime as follows
\begin{eqnarray}
d^2 (m_1,m_2) = \frac{\left( U(m_2) - U(m_1) \right) \left( V(m_1) - V(m_2) \right)}{ W(m_2) W(m_1)}.
\label{geo-lenght}
\end{eqnarray} 

\subsection{Without Islands}
\label{Sec: No Islands}

When there are no islands, the entropy of the Hawking radiation is given by the finite part of the EE of the matter as follows
\begin{eqnarray}
S_{ \rm \mathcal{R}} &=& \frac{c}{3} \log d(b_+, b_-)  \nonumber \\
                     &=& \frac{c}{6} \log  \left[ \frac{4} {\kappa^2}\frac{(b-r_h)(4 r_h \nu^2+3 b(\nu^2-1)) }{4 r_h^2 \nu^2} \cosh^2 \left(\kappa t_b \right) \right].
\label{S-R-CFT-no islands-2}
\end{eqnarray}
At early times, i.e. $t_b T \ll 1$, it grows quadratically in time
\begin{eqnarray}
S_{ \rm \mathcal{R}} =   \frac{c}{6} \left[  \log \left(\frac{(b-r_h)(4r_h \nu^2 + 3 b(\nu^2-1))}{r_h^2\kappa^2 \nu^2}\right) \right] + \frac{c}{6}\kappa^2 t_b^2 +\cdots  \ .
\label{S-R-CFT-no islands-early-times}
\end{eqnarray} 
The first term corresponds to the entanglement entropy of the radiation region in its initial state. However, it includes an undetermined constant that does not affect the overall growth behavior of the entropy. Moving on to the second term reflects the behavior of the entanglement entropy over time, particularly during the early stages. In this case, the entropy demonstrates a quadratic growth pattern, implying that its increase is proportional to the square of the elapsed time.
\par

On the other hand, at late times, i.e. $t_b T \gg 1$, it grows linearly in time
\begin{eqnarray}
S_{ \rm R} \simeq \frac{c}{3}    t_b \ .
\label{S-R-CFT-no islands-late-times}
 \end{eqnarray}

This expression indicates that the entanglement entropy of Hawking radiation exhibits a linear increase with time and diverges to infinity as the parameter \say{$t_b$} approaches infinity. This divergence results from the large separation between the two regions, $\mathcal{R}_-$ and $\mathcal{R}_+$, at later times. The infinitely large value of the entanglement entropy as $t_b$ approaches infinity signifies that the information regarding the initial-state matter that collapsed into the black hole, as well as the information about the particles that fell into it, cannot be recovered from the emitted Hawking radiation. Essentially, the contribution of the \say{no island} configuration to the entanglement entropy results in a non-unitary evolution of the black hole during evaporation. Now, we will demonstrate that the challenge to maintain the unitarity of quantum mechanics can be effectively addressed by introducing the concept of an \say{island} configuration. This configuration arises during the later stages of the black hole's evaporation process and provides a solution to the conflict.

\subsection{With Islands}
\label{Sec: With Islands}
The formula for the entropy of the two interval $\mathcal{A}$ and $\mathcal{B}$ is calculated by the formula as \cite{Casini:2005rm}
\begin{eqnarray}\label{EE-CFT-two disjoint interval-1}
S (\mathcal{A} \cup \mathcal{B}) = \frac{c}{3} \log \left( \frac{(b_1 - a_1) (b_1 - a_2) (b_2 - a_1) (b_2 - a_2)}{\epsilon^2 (a_2 - a_1) (b_2 - b_1)} \right) \ ,
\end{eqnarray} 
where $c$ is the central charge and $\epsilon$ is the UV cutoff.  After considering the finite part of \eqref{EE-CFT-two disjoint interval-1}, and noticing that our background is conformally flat, one has
\footnote{Notice that the metric \eqref{metric-black brane-Kruskal-1-new} is conformally flat and contain the warp factor $\mathcal{W}(r)$ introduced in \eqref{metric-black brane-Kruskal-1-new W(r)}.}
\bea
S_{\rm matter}^{\rm f} (\mathcal{R} \cup \mathcal{I}) = \frac{c}{3} \log \left( \frac{d(a_+,a_-) d(b_+,b_-) d(a_+,b_+) d(a_-,b_-)}{d(a_+,b_-) d(a_-,b_+)} \right).
\label{EE-CFT-two disjoint interval}
\eea 
Next, by applying the distance formulae, one obtains
\begin{equation*}
    S_{\rm matter}^{\rm f} (\mathcal{R} \cup \mathcal{I})  = \frac{c}{3} \Bigg[
\log \left( \frac{4\; \cosh{\left(\kappa t_a\right)} \cosh{\left(\kappa t_b\right)}}{\mathcal{W}(a)\mathcal{W}(b)} \right) +  \log \left( \frac{ \cosh \left( \frac{2 \pi}{\beta} (b - r^*(a)) \right) - \cosh \left(  \frac{2 \pi}{\beta} (t_a - t_b) \right)}{ \cosh \left( \frac{2 \pi}{\beta} (b - r^*(a)) \right) + \cosh \left(  \frac{2 \pi}{\beta} (t_a + t_b) \right) }\right)
\Bigg] \ . 
\end{equation*}
 
After adding the gravitational part, the generalized entropy becomes
\bea 
S_{\rm gen}  &=& \frac{\pi  R(a)}{ G_{N,r}} + S_{\rm matter}^{\rm f}  (\mathcal{R} \cup \mathcal{I})
\cr&&\cr
&=& \frac{\pi  R(a)}{G_{N,r}} + 
\frac{c}{3} \Bigg[
\log \left( \frac{4\; \cosh{\left(\kappa t_a\right)} \cosh{\left(\kappa t_b\right)}}{\mathcal{W}(a)\mathcal{W}(b)} \right)
\cr && \cr
&& + \log \left( \frac{ \cosh \left( \frac{2 \pi}{\beta} (r^*(b) - r^*(a)) \right) - \cosh \left(  \frac{2 \pi}{\beta} (t_a - t_b) \right)}{ \cosh \left( \frac{2 \pi}{\beta} (r^*(b) - r^*(a)) \right) + \cosh \left(  \frac{2 \pi}{\beta} (t_a + t_b) \right) }\right)
\Bigg]
\label{S-gen-CFT-islands}
\eea 
Here, the first term arises from the two-sided area of the island's boundaries. On the other hand, the second and third terms represent the contributions of the matter fields in the combined radiation and island regions. These terms collectively account for the overall entropy associated with the system. 

Considering the island's presence, we will examine the characteristics of the entanglement entropy during both the early and late stages of black hole evaporation.

\subsubsection{Early Times}

During the initial stages of black hole evaporation, the entanglement entropy of Hawking radiation is relatively low. Consequently, the island is expected to be located within the black hole. Furthermore, we assume that the boundaries of the radiation region are significantly distant from the event horizon of the black hole. Based on these considerations, we can utilize the following approximation \cite{Hashimoto:2020cas} as $r_h \ll b,\;\; t_a, t_b \ll 1/\kappa \ll r^*(b) - r^*(a)$. By the use of this approximation, the \eqref{S-gen-CFT-islands} can be written as 
\begin{eqnarray}
    S^{\text{early times}}_{\rm gen} &=& \frac{\pi  R(a)}{G_{N,r}} + 
\frac{c}{3} \left[ \kappa r^*(a)-\log \mathcal{W}(a) \right] + \frac{c}{6} \kappa^2 t_a^2 + \cdots \ .
\label{S-gen-CFT-islands}
\end{eqnarray}
To determine the locations of the island boundaries that yield the extremum of the generalized entropy, $S^{\text{early times}}_{\rm gen}$, we must minimize $S^{\text{early times}}_{\rm gen}$ as given in \eqref{S-gen-CFT-islands} concerning all possible $(a,  t_a)$ to get the positions of the island. The equations we have to solve is 
\begin{equation*}
    \frac{\partial S^{\text{early times}}_{\rm gen}}{\partial a}=0 \;\;\;\;\;;\;\;\;\;\;    \frac{\partial S^{\text{early times}}_{\rm gen}}{\partial t_a}=0 \ .
\end{equation*}
The equation does not possess any solution because the size of the island has been computed to be smaller than the Planck length, and we are neglecting all physics at the Planck scale. Additionally, selecting the extreme surface inside the Cauchy horizon is not permissible. The outcome suggests no real extremal point, confirming the absence of any nonvanishing "Quantum Extremal Surface" (QES) that could lead to the generalized entropy reaching an extremal value. This result aligns with our initial expectation that the island is not present during early times. Consequently, the entanglement entropy is solely determined by the contribution from radiation and exhibits a linear increase with time, consistent with \eqref{S-R-CFT-no islands-late-times}.

\subsubsection{At Late Times}

Let's examine the late-time behavior of the entropy of Hawking radiation. In the final stages of black hole evaporation, as the amount of radiation increases, the entanglement entropy contributed by the matter component also grows. Eventually, when the contribution of the second term, $ S_{\text{ matter}}^{\rm f}  (\mathcal{R} \cup \mathcal{I})$, in the formula for the calculation of $S_{\text{gen}}$ reaches the order of $\mathcal{O}(G_N^{-1})$. It becomes comparable to the first term in magnitude. At this point, a phase transition takes place. Consequently, Hawking radiation's fine-grained entropy starts to decrease, and the behavior of the entanglement entropy becomes limited by the principle of unitarity, aligning with our expectations.

\par Let's analyze the behavior of the entanglement entropy at late times. In this scenario, we adopt the following approximation, as described in \cite{Hashimoto:2020cas,Wang:2021woy}
\begin{equation*}
    t_a, t_b \gg r^*(b)-r^*(a) \gg \frac{1}{\kappa} \ ,
    \end{equation*}
    that leads to 
    \begin{eqnarray}
        \cosh{\kappa t_a} \simeq \frac{1}{2} e^{\kappa t_a} \;\;\;\;\;;\;\;\;\;\;  \cosh{\kappa t_b} \simeq \frac{1}{2} 
  e^{\kappa t_a}  
  \cr&&\cr
  \cosh{(\kappa(t_a+t_b))} \gg \cosh{(\kappa(r^*(b)-r^*(a)))} 
   \cr&&\cr
  \cosh{(\kappa(r^*(b)-r^*(a)))} \simeq \frac{1}{2}e^{(\kappa(r^*(b)-r^*(a)))} \ .
  \end{eqnarray}
  Then the expression 
  \begin{eqnarray}
     && \frac{c}{3}\log \left[\cosh{(\kappa t_a)}\cosh{(\kappa t_b)}  \frac{ \cosh \left( \kappa (r^*(b) - r^*(a)) \right) - \cosh \left(  \kappa (t_a - t_b) \right)}{ \cosh \left( \kappa (r^*(b) - r^*(a)) \right) + \cosh \left(  \kappa (t_a + t_b) \right) } \right]
     \cr && \cr
     && \;\;\;\;\;\;\;\;\;\;\;\;\;\;\;\;\;\;\;\;\;\;\;\;\;\;\;\;\;\; \to \frac{c}{3} \log \left[\cosh \left( \kappa (r^*(b) - r^*(a)) \right)- \cosh \left(  \kappa (t_a - t_b) \right)\right] \ .  \end{eqnarray}

      By analyzing the expression, it becomes evident that the maximal value is attained when the value of $\cosh{[\kappa(t_a-t_b)]}$ reaches its minimum, which occurs at $t_a=t_b$. Setting $t_a=t_b=t$ and substituting this value into the expression, we obtain an equation that is independent of time. This observation suggests that the entanglement entropy converges as time progresses, reaching a constant value at late times.
      
Now with all these approximations, we have 
\begin{eqnarray}
    S_{\rm gen} \simeq \frac{\pi R(a)}{G_{N,r}} - \frac{c}{3}\log\left\{\mathcal{W}(a) \mathcal{W}(b)\right\} +\frac{2c}{3} \kappa r^*(b) - \frac{2c}{3} e^{-\kappa(r^*(b)-r^*(a))}
\end{eqnarray}

We examine the scenario where the island is situated close to the event horizon, specifically at a distance denoted as $a = r_h + \epsilon + \mathcal{O}(\epsilon^2)$, where $\epsilon$ is a small parameter with $\epsilon \ll 1$. By employing perturbation techniques and solving the extremizing condition $\partial S_{\rm gen}/\partial a = 0$, we obtain the following results:

      \begin{equation}\label{Location of Island}
          a=r_h+\frac{4cG_{N,r}}{\pi (7 \nu^2-3)} + \frac{16 c^2 G^2_{N,r}(7\nu^2-15 \nu^4)}{3\pi^2 r_h^2(3+\nu^2)(7\nu^2-3)^2} + \cdots \ .
      \end{equation}

      The position of the island's boundary outside the event horizon is evident and consistent with Fig. \ref{fig: Penrose}, by substituting this location into \eqref{Location of Island}, we can obtain the actual fine-grained entropy of Hawking radiation as 

      \begin{equation}\label{S-R-CFT-with islands-late-times}
          S_{\rm gen} \simeq 2 S_{\rm BH} + \mathcal{O}(c G_{N,r}) \ .
      \end{equation}
      The dominant term in this expression corresponds to the Bekenstein-Hawking entropy of the black hole, as defined in \eqref{S-R-CFT-with islands-late-times}. It arises from incorporating the island's construction. Subsequent subleading and higher-order terms can be considered negligible compared to $S_{\rm BH}$.

\begin{figure}[ht]
	\begin{center}
		\includegraphics[scale=0.60]{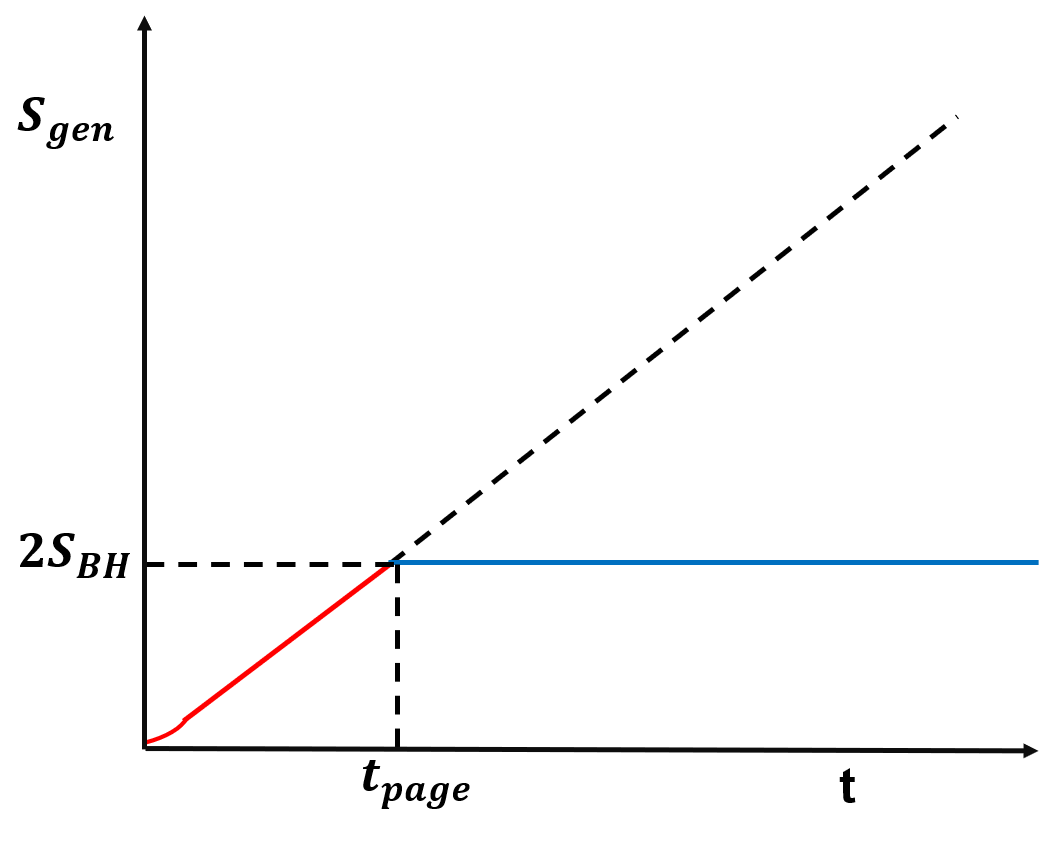}
	\end{center}
	\caption{
	The page curve illustrates the behavior of the wraped AdS black hole. The red line represents the entropy without the island, while the blue solid line depicts the saturation value of the entanglement entropy, taking the island into account.}
	\label{fig: ff}
\end{figure}

      \par
      Based on these findings, we can briefly discuss the behavior of generalized entropy. When there is no island in the early stages, the generalized entropy is primarily influenced by the matter's contribution and increases with time. However, as we reach later stages, the island's presence outside the event horizon becomes crucial for the generalized entropy to reach its minimal value. At this point, the generalized entropy becomes saturated and approaches a constant value, with the dominant term being the Bekenstein-Hawking entropy. As a result, we successfully replicated the Page curve, as depicted in Figure \ref{fig: ff}.

\section{Page time and scrambling time}
\label{Sec: Page time and scrambling time}
In this section, we will analyze the Page time and the scrambling time while exploring the influence of the parameter $\nu$ on the outcomes. Page time refers to the moment when the radiation entropy of the entire system reaches its maximum value. For an eternal black hole, its entropy remains unchanged after this time. However, its entropy decreases beyond this point for an evaporating black hole. The relationship between the black hole's lifetime and the Page time was discussed in a previous study \cite{Page}.

\par

The Page time can be estimated by identifying the point of intersection between the entropy curve without the island during early times and the entropy curve with the island during late times. Alternatively, we can determine the Page time by equating \eqref{S-R-CFT-no islands-late-times} and \eqref{S-R-CFT-with islands-late-times}. Let them be nearly equal during the late times and perform the calculation as follows
\begin{eqnarray}\label{Page time}
    t_{\rm page} &\simeq& \frac{6 S_{\rm BH}}{c \kappa} \nonumber \\
                &\simeq& \frac{3 S_{\rm BH} }{c \pi T} 
\end{eqnarray}
where $\kappa$ is surface gravity and $S_{\rm BH}$ is the black hole's Bekenstein-Hawking entropy.

\par

The scrambling time refers to the duration for the information that entered the black hole and subsequently emerged as Hawking radiation to be recoverable. In the context of the island prescription, the scrambling time aligns with the moment the information reaches the island. Suppose we dispatch a message from the radiation region $\mathcal{R}(r = b)$ at time $t = 0$; in that case, it will arrive at the island region $\mathcal{I}(r = a)$ precisely at time $t_0$. The scrambling time $t_{\rm scr}$ is 
\begin{eqnarray}
    t_{\rm scr} &=& t_0-t \nonumber \\
                &=& r^*(b)-r^*(a) \nonumber\\
                &\simeq& \frac{2 \nu}{\nu^2+3} \log S_{\rm BH} \nonumber \\
                 &\simeq& \frac{2 \nu}{\nu^2+3} \log \left\{\frac{\pi \nu r_h}{2 G_N}\right\} \ ,
\end{eqnarray}
here $r_h$ and $b$ have same magnitude. This expression indicates that the scrambling time is significantly smaller than the Page time and its behavior in the Warped AdS black hole parameters.
\section{Discussion}
\label{Sec: Discussion}

This paper investigates the black hole information paradox concerning Warped AdS black holes. Initially, we assume that a Warped AdS black hole exists in a pure quantum state. As the evaporation progresses, the amount of Hawking radiation produced tends to become infinite, leading to an infinite thermal radiation entropy. This exceeds the Bekenstein entropy bound and conflicts with the unitarity of quantum mechanics. However, for a real evaporating black hole, the information issue is fundamentally similar to that of an eternal black hole. In the final stage of the evaporation process, the black hole completely dissipates, transforming into Hawking radiation. Considering that it starts with the same initial condition, the fine-grained entropy of radiation should eventually decrease to zero. Nevertheless, the thermal radiation entropy remains much larger than the entropy bound. As a result, even in the case of eternal black holes, studying the information paradox remains highly relevant and significant.
\par
We have investigated the information problem concerning the well-known solutions to Einstein's equation, specifically the wrapped AdS black hole spacetime. Initially, no island formation occurs. This is attributed to the quantum entanglement entropy of the gravitational system, which is typically the minimum of two components in many scenarios: the island's area and the entanglement radiation within its wedge.
During the early stages, there isn't enough radiation generated, and the primary contribution to the entanglement entropy arises from the radiation itself, rendering the island unnecessary. However, as time progresses, the radiation becomes the dominant contributor, and the entanglement mainly stems from the area of the island, located within the stretched horizon.
By employing the island configuration, we determined the scrambling time, which aligns with the results obtained from the Hayden-Preskill protocol and the Page time. This entire evolution of the entanglement entropy can be visualized through the Page curve shown in Figure \ref{fig: ff}.
\par
Furthermore, our construction exclusively focused on zero or one-island scenarios. However, it is essential to note that various patterns of island formation can arise in general. Our derived Page curve observed a distinct inflection point occurring precisely at the Page time. If multiple islands form around the Page time, it will likely result in a smoother transition at the edge of the Page curve. By analyzing the configuration of islands, we addressed whether the black hole information is conserved in the Warped AdS black hole spacetime.

\section*{Acknowledgment}

I want to thank Farzad Omidi very much for the initial stage of this project and for having illuminating discussions during this work. We are also very grateful to Prasanta Tripathy for motivating me to work on the entanglement and for his blessings. I want to thank Amrendra Kumar for all Images.



\begin{appendices}

\section{Isometries}\label{Isometries of WAdS3}
This appendix will demonstrate that the familiar warped black holes are obtained as quotients of warped $AdS_3$ by applying a discrete subgroup $\Gamma$ of the isometry group, similar to how BTZ black holes are derived as quotients of $AdS_3$ \cite{Banados:1992wn,Banados:1992gq}. We can start with the local equivalence between warped black holes and warped $AdS_3$ arises when examining coordinate-invariant quantities. In the three-dimensional context, these quantities are constructed based on the Ricci tensor and its derivatives. The invariants constructed from the Ricci tensor are expressed as follows:

\begin{equation}
    \left[R,\;R_{\alpha \beta} R^{\alpha \beta},\;R_{\alpha \beta}R^{\alpha \gamma}R^\beta_\gamma \right] = \frac{6}{l^2} \left[-1,\;\frac{3-2\nu^2+\nu^4}{l^2},\;\frac{-9+9\nu^2-3\nu^4-\nu^6}{l^4}\right] \ .
\end{equation}
So, it is found that warped $AdS_3$ and the warped black hole solutions share identical values for these curvature invariants. We also have another invariant from the derivatives of the Ricci tensor
\begin{equation*}
    \nabla_{\alpha} R^{\beta \gamma} \nabla_{\gamma} R^\alpha_\beta = \frac{18 \nu^2\left(\nu^2-1\right)^2}{l^6} \ .
\end{equation*}

The observed concurrence between the coordinate invariant quantities of warped $AdS_3$ and the warped black holes strongly implies their local equivalence (for a more rigorous discussion, one can have a look on\cite{Sousa:2007ax}). We will now proceed to demonstrate this equivalence directly.

\par 

As the warped black holes are locally identified with warped $AdS_3$, they must result from quotienting the latter by a discrete subgroup $\Gamma$ of the isometries associated with $SL(2, \mathbb{R}) \times U(1)$, similar to the case of the BTZ black hole \cite{Banados:1992gq}. This section aims to determine the specific subgroup $\Gamma$. To achieve this, we identify points $\mathbf{P}$ in warped $AdS_3$ using a Killing vector $\xi$, which defines a one-parameter subgroup of the complete isometry group,
\begin{equation*}
    \mathbf{P} \sim e^{2\pi n \xi} \mathbf{P} , \;\;\;\;\;\;\;\;\;\;\; n=0,1,2,3,\cdots  \ .
\end{equation*}

The isometries associated with the different types of warped $AdS_3$ are:
\begin{itemize}
    \item The $SL(2, \mathbb{R})_L$ isometries are
    \begin{eqnarray}
    J_1 &=& \frac{2 \sinh{u}}{\cosh{\sigma}} \partial_\tau - 2\cosh{u} \partial_\sigma + 2 \tanh{\sigma} \sinh{u} \partial_u 
    \cr&&\cr
    J_2 &=& 2 \partial_u
    \cr&&\cr
    J_0 &=&  -\frac{2 \cosh{u}}{\cosh{\sigma}} \partial_\tau + 2\sinh{u} \partial_\sigma + 2 \tanh{\sigma} \cosh{u} \partial_u \ .
\end{eqnarray}
The isometries of the various types of warped $AdS_3$ satisfy the following algebraic relations:
\begin{equation*}
    [J_1, J_2] = 2 J_0 \;\;\;\;\;;\;\;\;\;\;[J_0, J_1] = -2 J_2\;\;\;\;\;;\;\;\;\;\;[J_0, J_2] = 2 J_1 \ .
\end{equation*}

\item The $SL(2, \mathbb{R})_R$ isometries are

 \begin{eqnarray}
    \Bar{J}_1 &=& 2 \sin{\tau} \tanh{\sigma} \partial_\tau - 2\cos{\tau} \partial_\sigma + \frac{2 \sin{\tau}}{\cosh{\sigma}}\partial_u 
    \cr&&\cr
    \Bar{J}_2 &=&  -2 \cos{\tau} \tanh{\sigma} \partial_\tau - 2\sin{\tau} \partial_\sigma - \frac{2 \cos{\tau}}{\cosh{\sigma}}\partial_u
    \cr&&\cr
    \Bar{J}_0 &=& 2 \partial_\tau \ .
\end{eqnarray}
The isometries of the various types of warped $AdS_3$ satisfy the following algebraic relations:
\begin{equation*}
    [\Bar{J}_1, \Bar{J}_2] = 2 \Bar{J}_0 \;\;\;\;\;;\;\;\;\;\;[\Bar{J}_0, \Bar{J}_1] = -2 \bar{J}_2\;\;\;\;\;;\;\;\;\;\;[\Bar{J}_0, \Bar{J}_2] = 2 \Bar{J}_1 \ .
\end{equation*}
\end{itemize}
In spacelike warped anti-de Sitter space, the preserved Killing vectors are generated by the $SL(2, \mathbb{R})_R$ group and $J_2$. These Killing vectors represent the symmetries that remain invariant under transformations and define the isometry group of spacelike warped $AdS_3$. So, the isometry group of spacelike warped AdS is $U(1)_L \times SL(2, \mathbb{R})_R$. The most general expression for the Killing vector, along which we can perform a quotient, is as follows: 
\begin{equation*}
    \beta_2 J_2 + \alpha_2 \Bar{J}_2 + \alpha_0 \Bar{J}_0 \ .
\end{equation*}
There are three types of one-parameter subgroups generated by the isometry group $U(1)_L \times SL(2, \mathbb{R})_R$ in spacelike warped AdS. These subgroups correspond to different symmetries and transformations that preserve the geometry of the spacetime as

\begin{eqnarray}
    \eta_a &:& \beta_2 J_2 +\alpha_0 \Bar{J}_0 \nonumber \\
    \eta_b &:& \beta_2 J_2 +\alpha_2 \Bar{J}_0 \nonumber \\
     \eta_c &:& \beta_2 J_2 + \Bar{J}_0 + \Bar{J}_2 \ .
\end{eqnarray}

According to the classification in \cite{Banados:1992gq}, these correspond to $I_a$, $I_b$, and $II_a$ (the first form), respectively. The norms of the aforementioned generators are 

\begin{eqnarray}
    \eta^2_a &=& \frac{12 l^2 (\nu^2-1)}{(\nu^2+3)^2}\left\{\alpha_0 \sinh{\sigma}\right\}^2+\frac{32 l^2 \nu^2 \beta_2}{(\nu^2+3)^2}\left\{\alpha_0 \sinh{\sigma}\right\} - \frac{4 l^2 \alpha_0^2}{\nu^2+3}+\frac{16 l^2 \nu^2 \beta_2^2}{(\nu^2+3)^2} \nonumber \\
    \eta^2_b &=& \frac{12 l^2 (\nu^2-1)}{(\nu^2+3)^2}\left\{\alpha_2 \cos{\tau} \cosh{\sigma}\right\}^2 - \frac{32 l^2 \nu^2 \beta_2}{(\nu^2+3)^2}\left\{\alpha_2 \cos{\tau} \cosh{\sigma}\right\} + \frac{4 l^2 \alpha_2^2}{\nu^2+3}+\frac{16 l^2 \nu^2 \beta_2^2}{(\nu^2+3)^2} \nonumber \\
    \eta^2_c &=&\frac{12 l^2 (\nu^2-1)}{(\nu^2+3)^2}\left\{\sinh{\sigma}- \cos{\tau} \cosh{\sigma}\right\}^2 + \frac{32 l^2 \nu^2 \beta_2}{(\nu^2+3)^2}\left\{\sinh{\sigma} - \cos{\tau} \cosh{\sigma}\right\} +\frac{16 l^2 \nu^2 \beta_2^2}{(\nu^2+3)^2} \ . \nonumber
\end{eqnarray}

\par

Identifying points along $\partial_\phi$ direction with the identification $\phi\sim \phi+2 \pi$. After representing the $\partial_\phi$ Killing vector using the original warped anti-de Sitter coordinates, we proceed to discretely quotient along the isometry
\begin{equation}
    \xi = \frac{1}{2\pi}\partial_\phi=\frac{\nu^2+3}{16 \pi} \left[\left(r_++r_--\frac{\sqrt{(\nu^2+3)r_+r_-}}{\nu}\right)J_2 - (r_+-r_-)\Bar{J}_2\right] \ ,
\end{equation}
where $J_2 \in U(1)_L$ and $\Bar{J}_2 \in SL(2,\mathbb{R})_R$. Now, one can define the right and left moving temperatures as 
\begin{eqnarray}
    T_R &=& \frac{(\nu^2+3)(r_+-r_-)}{8 \pi l} \nonumber \\
    T_L &=& \frac{(\nu^2+3)}{8 \pi l} \left(r_++r_--\frac{\sqrt{(\nu^2+3)r_+r_-}}{\nu}\right) \ .
\end{eqnarray}
In the terms of $T_R$ and $T_L$ we have the Hawking temperature($T_H)$ as 
\begin{eqnarray}
    T_H &=& \frac{\nu^2+3}{4 \pi \nu l} \frac{T_R}{T_R+T_L} \nonumber \\
        &=& \frac{\nu^2+3}{4 \pi l} \frac{(r_+-r_-)}{(2 \nu r_+ - \sqrt{(\nu^2 +3)r_+r_-})} \ .
\end{eqnarray}

\end{appendices}




\end{document}